# Observation of Nanoscale Opto-Mechanical Molecular Damping; Origin of Spectroscopic Contrast in Photo Induced Force Microscopy

Mohammad A. Almajhadi[1], Syed Mohammad Ashab Uddin[1], and H. Kumar Wickramasinghe*[1]

[1] Department of Electrical Engineering and Computer Sciences, University of California, Irvine, Irvine 92697, CA, USA. Correspondence and requests for materials should be addressed to H. Kumar Wickramasinghe (email: hkwick@uci.edu)

**Abstract:**

We experimentally investigated the contrast mechanism of infrared photoinduced force microscopy (PiFM) for recording vibrational resonances. Extensive experiments have demonstrated that spectroscopic contrast in PiFM is mediated by *opto-mechanical damping of the cantilever oscillation* as the optical wavelength is scanned through optical resonance. To our knowledge, this is the first time opto-mechanical damping has been observed in the AFM. We hypothesize that this damping force is a consequence of the dissipative interaction between the sample and the vibrating tip; the modulated light source in PiFM modulates the effective damping constant of the $2^{nd}$ eigenmode of the cantilever which in turn generate side-band signals producing the PiFM signal at the $1^{st}$ eigenmode. A series of experiments have eliminated other mechanisms of contrast. By tracking the frequency shift of the PiFM signal at the $1^{st}$ cantilever eigenmode as the excitation wavenumber is tuned through a mid-infrared absorption band, we showed that the near-field optical interaction is attractive. By using a vibrating piezoelectric crystal to mimic sample thermal expansion in a PiFM operating in mixing mode, we determined that the minimum thermal expansion our system can detect is 30 pm limited by system noise. We have confirmed that van der Waal mediated thermal-expansion forces have negligible effect on PiFM signals by detecting the resonant response of a 4-methylbenzenethiol mono molecular layer deposited on template-stripped gold, where thermal expansion was expected to be < 3 pm, i.e., 10 times lower than our system noise level. Finally, the basic theory for dissipative tip-sample interactions was introduced to model the photoinduced opto-mechanical damping. Theoretical simulations are in excellent agreement with experiment.

I. Introduction:

The integration of atomic force microscopy (AFM) with focused lasers has enabled nano-chemical imaging and spectroscopy with spatial resolution well beyond the diffraction limit. One classic example is apertureless near-field scanning optical microscopy (a-NSOM or sSNOM)[1–4]. In this method, the enhanced optical field of the scanned AFM probe is perturbed by the local near-field generated by the excited sample and the scattered near-field (amplitude and phase) is detected in the far-field using an interferometer to record the image. Photothermal induced resonance (PTIR)[5,6] and peak force infrared (PFIR)[7] are two examples for characterizing sample chemical properties based on AFM. In these techniques, the sample thermal expansion induced by optical absorption is detected using an AFM tip in contact mode. Despite the success of these two techniques, imaging soft samples with AFM in contact mode is likely to be challenging due to possibilities of sample damage. An alternative, noninvasive microscopy and spectroscopy technique that has emerged recently is photoinduced force microscopy (PiFM)[8] (Fig. 1). In this method, the tip-sample optical interaction is measured with the AFM operating in non-contact mode. The topography is recorded using the 2$^{nd}$ mechanical eigenmode of the cantilever at $f_2$. A quantum cascade laser (QCL) is amplitude modulated at $f_m$ (where $f_m = f_2 - f_1$) and focused on the tip end, and the opto-mechanical response is measured at the 1$^{st}$ mechanical eigenmode at $f_1$. Many applications of PiFM have emerged. Near-field electromagnetic field characterization[9–14], nonlinear optical measurements such as Raman[15] spectroscopy and stimulated Raman spectroscopy[16,17], time-resolved pump-probe microscopy[18], organic solar cells studies[19], optical phonon polariton imaging and nanoscale chemical imaging in the mid-infrared[20] are but a few examples. While the dipole-dipole force model provides excellent agreement with the electromagnetic near field measurements in the visible[14] and with mid-infrared plasmonic resonance spectra[21], extending this model to infrared vibrational resonances causes discrepancies between experiment and theory[20,22,23]. In particular, the dipole-dipole force model predicts a dispersive spectral response (or more accurately a combination of dispersive and dissipative responses), while the experimental results show a purely dissipative response. Three alternative proposals for explaining PiFM spectroscopic contrast in the infrared have been proposed to address this discrepancy. They are (1) detecting photothermal expansion using short range repulsive forces acting on the AFM cantilever/tip[24] in contact mode (2) detecting photoacoustic pressure waves generated at the sample surface resulting in long range repulsive forces acting on the cantilever/tip (3) detecting van der Waal mediated force modulation caused by sample thermal expansion[25].

In this paper, we report on a series of experiments aimed at unravelling the origin of PiFM spectroscopic contrast in the infrared. Our experimental findings support the hypothesis that the spectroscopic contrast in PiFM is mediated by *opto-mechanical damping of the cantilever oscillation* as the optical wavelength is scanned through optical resonance. We hypothesize that this damping force is caused by the excited sample molecules creating a dissipative force on the vibrating tip. We show that this contrast mechanism provides an excellent match with the experimental results. The theory can be extended to the single monolayer detection limit (see section IV).

Our paper is organized as follows. In section II, we study experimentally the repulsive/attractive nature of the optical forces exerted on the cantilever as a function of optical frequency by tracking its first eigenmode at $f_1$ – the channel typically used to record PiFM signals. In section III, the thermal expansion in photothermal microscopy is mimicked using a vibrating piezoelectric PZT crystal, and the minimum detectable thermal expansion for our setup is determined. In section IV, we present the first results of PiFM recorded spectra from a 4-methylbenzenethiol (4-MBT) mono molecular layer deposited on template-stripped gold (TSG) sample and show that the signal could not be explained by thermal expansion. Section V is devoted to a detailed experimental study aimed at identifying the true nature of the mixing signal in PiFM. Section VI introduces a theory for PiFM contrast based on opto-mechanical damping and compares experiment with theory. In section VII we discuss our results in the context of prior work. Finally, in section VIII we present some brief concluding remarks.

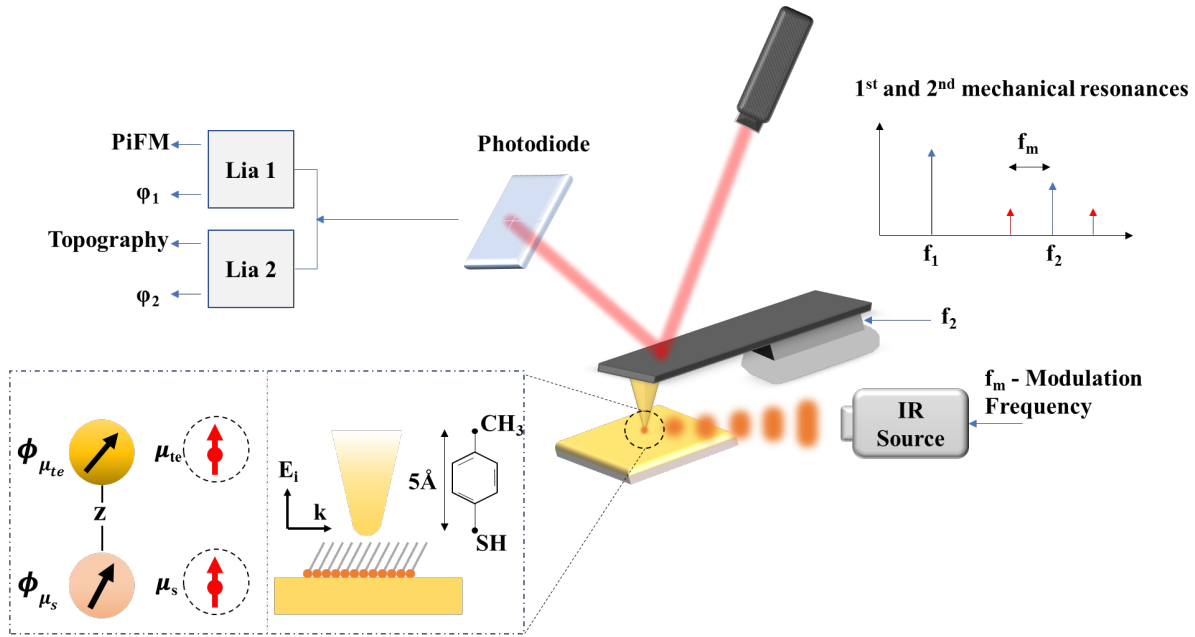

**Fig. 1** Schematic of IR PiFM experiment. The cantilever was mechanically vibrated at its 2$^{nd}$ mechanical eigenmode $f_2$, so that peak-peak oscillation was 6 nm. Lock-in amplifier and feedback laser position

sensitive detector (PSD) are used to stabilize the cantilever nanometers from sample surface. The IR source was electrically triggered at $f_m = f_2 - f_1$, where $f_1$ is the 1st mechanical eigenmode of the cantilever. The incident infrared pulse was p-polarized (along the tip axis) and focused to 20-um-diameter spot. The topography and the PiFM signals are simultaneously recorded at $f_2$ and $f_1$ respectively. The image was generated via raster-scanning the sample under the tip.

II. Frequency Tracking - Distinguishing Between Repulsive and Attractive Optical Forces

Force gradients acting on an AFM tip shifts its dynamic stiffness (k) and resonance frequency f. Frequency shifts at the second eigen mode $f_2$ due to optical force gradients are too small and cannot be detected due to the very high dynamic stiffness constant $k_2$ at $f_2$ ($k_2 = 39.31k_1$[26] where $k_1 = 9$ N/m). However, we *are* able to control the tip-sample gap using the second eigen mode at $f_2$ and measure frequency shifts at $f_1$ where the dynamic stiffness of the cantilever $k_1$ is much lower.

Fig. 2 shows results from a 60 nm thick polystyrene (PS) film on gold (Au) substrate. We record the frequency shift of the cantilever at $f_1$ while the tip sample gap is controlled using the second eigen mode at $f_2$. We plot the frequency shift at $f_1$ as we scan the optical excitation through the PS resonance. When sample is excited on resonance, the frequency of the 1st eigenmode shows a maximum shift toward lower frequency values, relative to off resonance excitation, revealing the *attractive nature* of the optical force. Previous works[27] have also come to the same conclusion, where the frequency shift was measured relative to the free oscillation amplitude (i.e. 3 μm away from the sample surface); however, in those experiments, non-optical van der Waal (vdW) effects were not subtracted. In our experiments, we automatically eliminate any vdW force gradient effects by measuring the frequency shift at $f_1$ as we scan through optical resonance while the tip is engaged. The observed frequency shifts were in the range of 500 Hz.

Short-range thermal-expansion forces in contact mode acting on the cantilever should manifest as two major features in the AFM cantilever dynamics: the thermal expansion force causes frequency of the eigenmode to shift to higher frequency values and the signal strength decays monotonically and rapidly as the cantilever is retracted from sample surface away from the repulsive regime. None of these features are observed in our experiments. In addition, as we shall see later, Fig. 6 (b) shows that PiFM signals are measurable in non-contact up to 18 nm from hard contact regime (shaded area). We conclude from all these measurements that short-range (repulsive) thermal-expansion forces are clearly not relevant in PiFM.

It is well known that energy absorbed at the surface of a sample can generate acoustic pressure waves in the surrounding gas – photoacoustics[28]. For our system, the pressure waves will have a wavelength ranging from 263 µm to 1.2 mm (corresponding to $f_m$ = 1.3 MHz and $f_1$ 250 kHz respectively). During one cycle of cantilever oscillation the change in near field photoacoustic force, i.e., acoustic force gradient, acting on the cantilever should be much smaller than the near field optical force gradient acting on the tip. Photoacoustics generated by the 20 µm IR spot on the sample could still exert a global repulsive force on the cantilever. We were indeed able to detect a global photoacoustic effect originating from the focused infrared beam for relatively thick samples (> 100 nm) (see Supplementary 1 Fig. 1) but only when the tip is retracted a few μm from sample when the much larger optical forces become negligible. We also observed that the global photoacoustic signal disappears when the system is operated in a vacuum of 0.3 torr. Based on these considerations, we conclude that the near field repulsive force due to gas photoacoustics will have minimal effect on the overall near field PiFM signal in our measurements. Our experimental observations in this section have refuted the proposals that gas photoacoustic forces or short-range thermal expansion forces play any significant role in PiFM contrast, at least in the regime that we have investigated – i.e. organic samples with thicknesses less than 60 nm. We will therefore no longer consider photothermal expansion or gas photoacoustics as potential contributors to PiFM contrast in later sections of this paper.

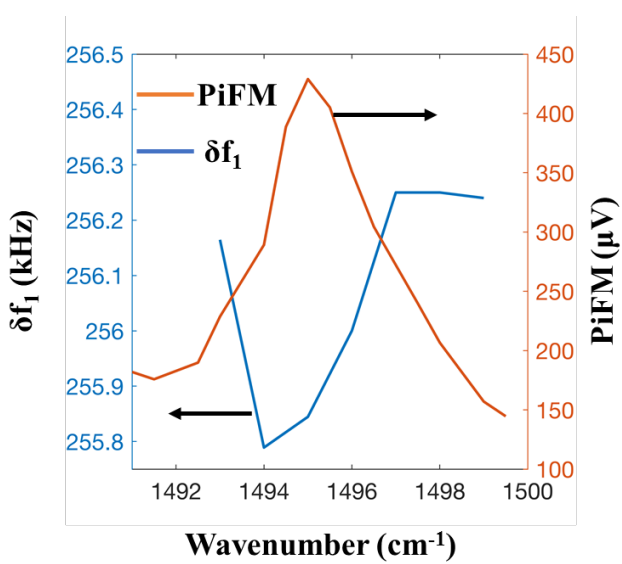

**Fig. 2** Shift of the cantilever resonance frequency at $f_1$ (blue line) across PS absorption band at 1495 cm$^{-1}$ (Orange line). The sample is 60 nm thick PS film on Au substrate. Input average power was 1 mW focused to 20-um-diameter spot.

III. Piezo Vibration Experiments and PiFM Sensitivity to Thermal Expansion:

In section II, we showed that thermal expansion and photoacoustics (short- and long-range repulsive forces respectively) do not play a significant role in our PiFM setup. This leaves us with thermally modulated vdW forces ($F_{th}^{vdW}$); i.e. thermal expansion modulates and amplifies the vdW force which in turn acts on the AFM tip and consequently generates the PiFM signal. The modulated $F_{th}^{vdW}$ are long range attractive forces. In this section, we mimic thermal expansion in our setup by vibrating a mirrored PZT crystal – which in turn modulates the vdW force. Our PZT crystal was independently calibrated using a heterodyne laser interferometer (see Supplement 2 Fig. 2). Experiments were carried out to determining the smallest detectable thermal expansion in our PiFM.

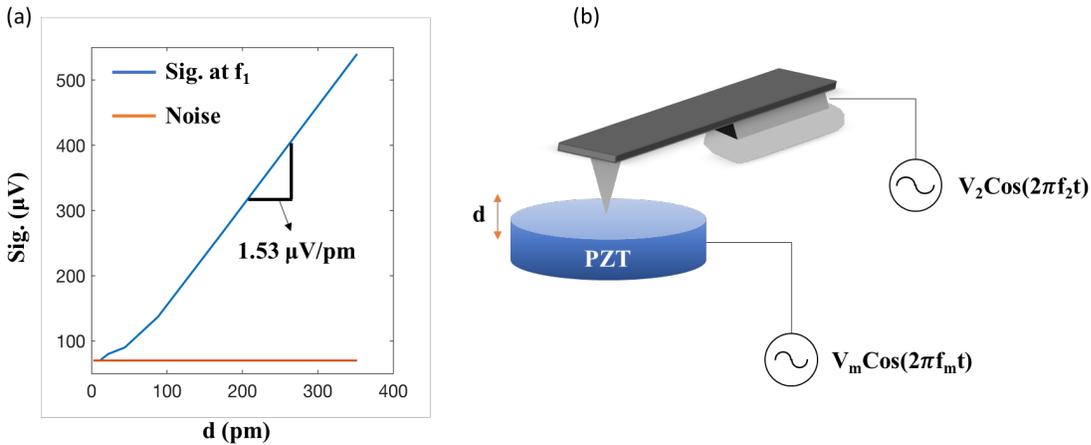

**Fig. 3** Non-contact AFM sensitivity to thermal expansion. **a** Cantilever response (blue solid line) measured at $f_1$ as a function of the PZT displacement amplitude. The noise level (50 uV) of the system is indicated by the orange line. Tip-sample gap was controlled by $f_2$ and the PZT was driven at $f_m = f_2 - f_1$. **b** depicts the experimental setup.

Figure 3b depicts our experimental set up. Gold coated PZT was vibrated at $f_m = f_2 - f_1$. The modulated vdW at $f_m$ mixes with $f_2$ to generate a signal at $f_1$ due to nonlinear tip-sample interactions. We plotted the sensitivity (S) defined as the ratio of the measured signal (mV) to the piezo displacement (pm) and compared it with the noise level of the PiFM to determine the minimum detectable thermal expansion. Results in Fig. 3a show a linear relationship between PZT displacement and the observed signal, with S = 1.53 µV/pm (Fig. 3a will be used later to estimate the thermal expansion contribution to our PiFM signal in our monolayer experiments). Since the noise level measured at $f_1$ is about 50 µV for 5 ms integration time, thermal expansion below 32 pm will not be detectable in our system. Tip-enhanced thermal expansion for a 60 nm PS film on silicon substrate ( excited at 1452 cm$^{-1}$ with 5 mW average power ) has been previously calculated to be about 30 pm[25], which is already below our noise level. That study shows that thermal signals generated by monolayer samples with typical thermal expansion of a few pm would be barely detectable. In the following section, the response of a 4-MBT monolayer on TSG was measured with a signal to noise ratio of 100, further confirming that our PiFM contrast cannot be thermal or $F_{th}^{vdW}$ in origin.

IV.     Monolayer PiFM Experiments

To demonstrate PiFM sensitivity to optical forces generated by molecular vibrational resonances of monolayer samples, 4-methylbenzenethiol (4-MBT), self-assembled monolayer (SAM) solution was prepared and TSG sample was immersed and left overnight in solution. The TSG is expected to be completely covered by a 4-MBT monolayer. Gold islands are generated by sonicating the TSG in ethanol until gold starts lifting off. Figure 4c shows the topography of the sample. The thickness of the monolayer is less than 5Å. The sample was excited with p-polarized light using quantum cascade laser (QCL). Measured average power was 0.5 mW. The diameter of the focal spot is 20 μm, with incident angle of 30° measured from sample surface. Tip-sample distance was controlled at $f_2$ with dithering amplitude of 6 nm peak to peak. Set point was adjusted such that average tip-sample distance is approximately 7 nm (refer to Fig. 6b). Thus, minimum tip-sample distance will be around 4 nm. QCL repetition rate was tuned such that the lower sideband $f_2 - f_m$ coincided with $f_1$. The PiFM signal is enhanced by the quality factor ($Q_1$) at $f_1$, which is about 100. In addition to the mechanical enhancement, the Silicon cantilever/tip was coated with 60-nm-thick gold to locally enhance the electromagnetic field.

Figure 4b shows the absorption spectrum of 4-MBT which is centered at 1495 cm$^{-1}$, with full width at half max (FWHM) of about 4 cm$^{-1}$. This sharp absorption band is a typical signature of benzene ring mode. Fig. 4c-d, respectively, show simultaneously recorded topography and PiFM images; Fig. 4e is PiFM image when 4-MBT is excited off resonance; it shows that the signal observed in Fig. 4b is not due to any possible cross talk between topography and PiFM. Under similar experimental conditions, thermal expansion of a monolayer has been calculated numerically to be < 3 pm with a temperature increase of < 6 degrees [29,30]. Also, we have shown in the previous section that minimum detectable thermal expansion is 32 pm – limited by our system noise level. Because the maximum measured signal for 4-MBT is about 600 μV, and since S = 1.53 μV/pm, the mono-molecular layer must expand 392 pm to generate our signal – almost 100% of its initial thickness! Such an expansion would imply heating the molecular layer by several 100s of degrees. We conclude that the observed PiFM signal clearly could not originate from thermal expansion. The findings also support the fact that any $F_{th}^{vdW}$ effect on the PiFM signal is negligible. In section V, experiments were performed to study the effect of vibrational resonances on the cantilever dynamics and to unravel the actual contrast mechanism in IR-PiFM.

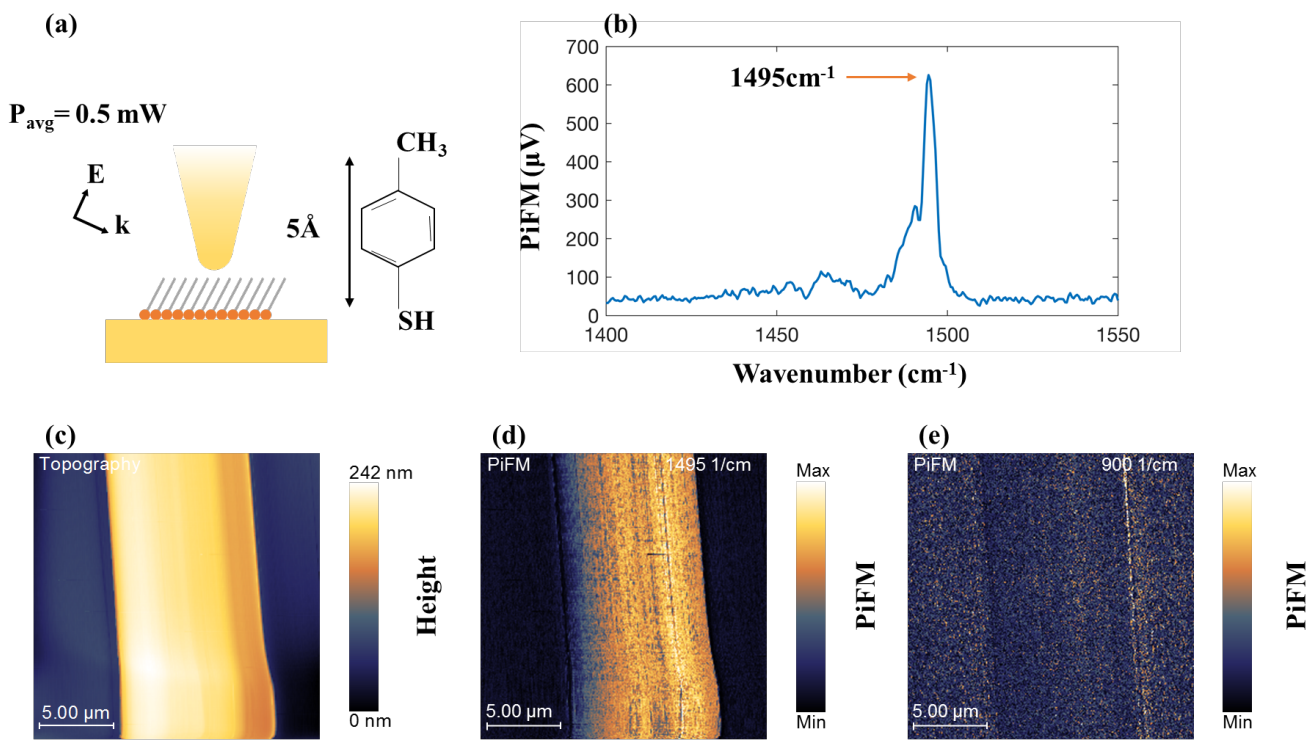

**Fig. 4 a** 4-MBT adsorbed on template-stripped gold. Excitation average power is 0.5 mW focused to 20-um-diameter spot, and the incident angle is 30º. **b** point spectrum of 4-MBT showing resonance at 1495 cm$^{-1}$. **c – e** are topography of gold island, PiFM image at 1495 cm$^{-1}$, and PiFM image off resonance respectivily.

V. Photoinduced Mechanical Dissipation (opto-mechanical damping):

The sidebands in PiFM can originate from either amplitude modulation or frequency modulation of the 2$^{nd}$ resonance $f_2$ of the cantilever. In one analysis, it was considered to originate from a frequency modulation of $f_2$ resonance[31]; a change in tip-sample interaction force (force gradient) leads to change in the effective spring constant, which intern shifts $f_2$ at the chopping frequency $f_m$. Amplitude modulation of $f_2$ is another way to generate sidebands. The excited molecule interacting with the tip exerts a damping force in the tip leading to a change in the cantilever effective mechanical damping constant, which in turn amplitude modulates $f_2$ at $f_m$. We conducted a series of experiments to determine whether the PiFM signal detected at $f_1$ is due to AM or FM modulation of cantilever second eigen mode.

resonance. In our experiments, the cantilever was first excited at a frequency slightly higher than $f_2$ ($f_{2R}$) and then excited at a frequency slightly lower than $f_2$ ($f_{2L}$). The sample was 60 nm PMMA on glass. The laser was tuned to a PS resonance and modulated at $f_m$. When we recorded the phase of the PiFM signal at $f_1$, we discovered that the $f_1$ signal had the *same phase* for both $f_{2R}$ and $f_{2L}$ experiments indicating that our PiFM signal contrast was originating from AM rather than FM modulation of the cantilever second eigen mode. We then performed another series of experiments to confirm these findings. The cantilever was mechanically excited at $f_{2R}$. The tip was approached and engaged with the sample. The feedback loop was opened, and the laser wavelength was rapidly swept across PMMA absorption band centered at 1733 cm$^{-1}$; the oscillation amplitude and phase $A_{2R}$, $\varphi_{2R}$ at $f_{2R}$ were recorded and compared with the point spectrum taken earlier for the same sample but with the control loop closed. The acquisition time needed to be fast enough to minimize thermal drift during data acquisition. In addition, laser modulation frequency $f_m$ was set at 1 MHz so that it did not excite any cantilever eigenmodes – i.e. in these studies we can consider the laser to be behaving essentially as a cw source of energy. Our experiments revealed – to our surprise – that the mechanical oscillation amplitude of the cantilever $A_{2R}$ *was damped as the laser was scanned through the PMMA resonance*! (Fig. 5e). The experiment was repeated again with excitation frequency at $f_{2L}$ with exactly the same result (Fig. 5f). Figure. 5a and 5d show the expected phase and amplitude response of two harmonic oscillators with different quality factors. If photoinduced force was dissipative, the predicted phase behavior for $f_{2R}$ and $f_{2L}$ is shown as a transition from a-a' and b-b'. As mentioned, $A_2$ *decreases* in both cases as shown in Fig. 5e and 5f, in contrast to what is expected from a conservative force. The change in $Q_2$ is evident from Fig. 5e and 5f, and it tracks the change in the point spectrum – similar to what was shown earlier. Another piece of evidence that demonstrates the dissipative nature of photoinduced force is shown in the phase measurements of Fig. 5 b–c. The phase of $f_2$ ($\varphi_2$) was recorded while the optical wavenumber was rapidly swept through resonance. According to Fig. 5 (a), we should expect to observe a decrease in the phase for $f_{2R}$ and an increase at $f_{2L}$ as verified in Fig. 5b and (c). The change $\varphi_R$ is 3° at $f_{2R}$ and the change $\varphi_L$ is 1° at $f_{2L}$. We note that the phase measurements and point spectrum were simultaneously recorded for the 60 nm PMMA film on glass.

Based on all our experiments, we conclude that a change in the *effective mechanical damping* constant of the cantilever rather than a change in its effective spring constant is the dominant contrast mechanism in PiFM. The intensity modulated excitation source modulates the mechanical damping at $f_m$ which in turn generates the mixing signal measured at $f_1 = f_2 - f_m$.

In our experiments, $f_2 \sim 1.6$ MHz and $Q_2 \sim 600$. Then, the intrinsic damping constant $m_2 \gamma_2 = \frac{k_2}{\omega_2 Q_2}$ (where $m_2$ is the effective mass at $f_2$), is approximately 58nN/m/s. From Fig. 5e and 5f we see that the cantilever oscillation is reduced by approximately 35% when the tip is stabilized 7nm from the sample surface and the optical wavelength is tuned to the molecular resonance. We conclude that the opto-mechanical damping constant is 0.35 x 58 nN/m/s or 20 nN/m/s at molecular resonance (see equation 6). The opto-mechanical damping constant reaches its maximum value when the molecule is driven at its optical resonance (i.e. at maximum optical polarization of the molecule).

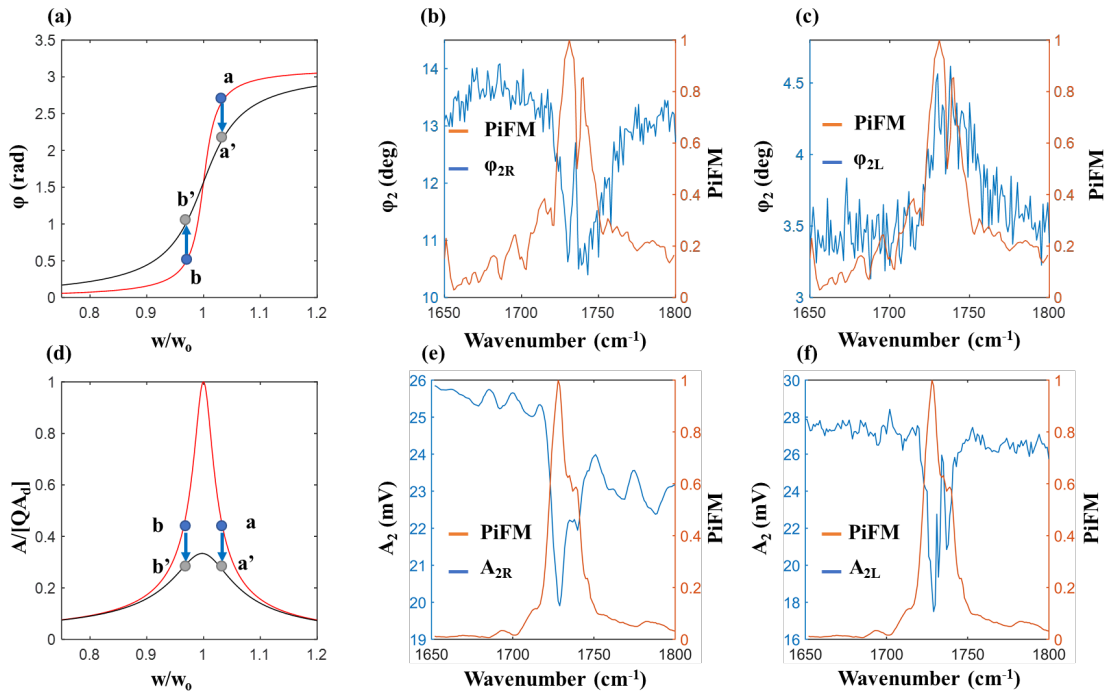

**Fig.5** Opto-mechanical damping. **a** and **d** are phase and amplitude of high Q (red) and low Q (black) harmonic oscillator. **b** and **c** are change in the phase of the 2$^{nd}$ mechanical mode ($\varphi_2$) (blue line) across

PMMA absorption band (orange line) centered at 1733 cm$^{-1}$ for right excitation and left excitation respectively. Change in the amplitude of the 2$^{nd}$ mechanical mode (A$_2$) (blue line) across PMMA absorption band (orange line) for right excitation **e** and left excitation **f**. Sample is 60 nm thick PMMA on glass. The phase measurement was conducted using standard PiFM mixing mode, while amplitude measurement was conducted with the feedback loop open and the wavenumber scanned rapidly through resonance as explained in the text.

VI. Theory

In PiFM, dissipation occurs due to damping introduced by the sample molecules interacting with the vibrating tip when the sample molecule is irradiated with photon energy corresponding to one of its molecular vibrational modes. We can model the change in the mechanical loss (Fig. 5) as follow[32]. The averaged power delivered to the cantilever excited at $\omega_c$ with driving amplitude $A_d$ is given by

$$<P_d> = \frac{1}{2} A A_d k \omega_c \sin(\phi) \quad (1)$$

where A and $A_d$ are oscillation amplitude of the cantilever and driving signal respectively. k and Φ are stiffness of the cantilever and the phase difference between driving signal and cantilever oscillation respectively. The loss $P_c$ due to intrinsic mechanical dissipation in the cantilever with quality factor $Q_c$ and resonance frequency $\omega_c$ is given by

$$<P_c> = \frac{1}{2Q_c} k A^2 \omega_c \quad (2)$$

If $<P_t>$ is the averaged total optical power supplied by the tip-sample cavity field for opto-mechanical damping of the cantilever, from power balance, we can write

$$<P_t> = <P_d> - <P_c> \quad (3)$$

Combining Eq. 1, 2 and 3 gives

$$<P_t> = \frac{kA^2\omega_c}{2Q_c}\left(\frac{Q_c A_d \sin(\emptyset)}{A} - 1\right) \quad (4)$$

$$A = Q_c A_d \sin(\emptyset)\left(\frac{<P_t>}{<P_c>} + 1\right)^{-1} \quad (5)$$

Since dissipated mechanical power is related to damping constant $\gamma$ through $P = 0.5\ m\gamma\omega_c^2 A^2$ (where m is the effective mass of the cantilever), we can equivalently write equation (5) in terms of the opto-mechanical damping constant $\gamma_o$ and the cantilever intrinsic damping constant $\gamma_c$.

$$A = Q_c A_d \sin(\emptyset)\left(\frac{\gamma_o}{\gamma_c} + 1\right)^{-1} \quad (6)$$

Equations (5) and (6) show that the cantilever oscillation amplitude decreases when the opto-mechanical damping constant $\gamma_o$ or the optically mediated tip-sample power dissipation increases – both will reach a maximum at optical resonance. Modulating the incident light intensity at $f_m$ ($f_m = f_2 - f_1$), produces maximum sideband oscillation amplitude and therefore, maximum signal at $f_1$ when the sample is driven at one of its vibrational resonances. In addition, in order to get maximum sensitivity for detecting molecular resonance, we need to choose an AFM setup with the lowest mechanical loss. S/N will be greatly improved by working even in a rough vacuum where air damping would be significantly minimized.

The z component of the optical force $F_{tz}$ acting on the tip with effective dipole moment $\mu_{te}$ can be written as[33]

$$F_{tz} = \mu_{te} \frac{dE_{tz}}{dz} \quad (7)$$

or

$$F_{tz}\, dz = \mu_{te} dE_{tz} = \alpha_{te}\, E_{tz} dE_{tz} \qquad (8)$$

Where $\alpha_{te}$ is the effective polarizability of the tip

$$F_{tz} \frac{dz}{dt} = \alpha_{te}\, E_{tz} \frac{dE_{tz}}{dt} \qquad (9)$$

The time averaged mechanical power $P_t$ dissipated due to optical forces acting on tip is

$$P_t = <F_{tz} \frac{dz}{dt}> = 0.5 Re\,[\alpha_{te}^* E_{tz}^* i\omega_0 E_{tz}] = Im\,[\alpha_{te}^* E_{tz}^* \omega_0 E_{tz}] \qquad (10)$$

where $\omega_0$ is the optical frequency and $E_{tz}$ the z-component of the electric field at tip

where

$$E_{tz} = \left(1 + \frac{\alpha_s}{2\pi(d+2a)^3}\right) E_i \qquad (11)$$

and $\alpha_{te}$ is given by[34]

$$\alpha_{te} \sim \alpha_t + \frac{\alpha_t \alpha_t \beta_s}{16\pi(d+2a)^3} \qquad (12)$$

$\alpha_t$ and $\alpha_s$ are the polarizabilities of tip and sample respectively, d is tip-sample distance, a is tip radius and $E_i$ is incident field

To arrive at equation (12) we made the approximation $\left[\frac{\alpha_t}{1-\frac{\alpha_t \beta_s}{16\pi(d+2a)^3}}\right] \sim \left[\alpha_t + \frac{\alpha_t \alpha_t \beta_s}{16\pi(d+2a)^3}\right]$

Using equations 10-12, the total optical power $P_t$ supplied by the tip-sample cavity field for mechanical damping of the cantilever can be written as

$$<P_t> \sim \frac{1}{2} Im \alpha_t^* E_i^2 + \frac{1}{2} Im\left(\frac{\omega_0 \alpha_t^* \alpha_s^* E_i^2}{2\pi\,(d+2a)^3}\right) + \frac{1}{2} Im\left(\frac{\omega_0 \alpha_t^* \alpha_s E_i^2}{2\pi\,(d+2a)^3}\right) + \frac{1}{2} Im\left(\frac{\omega_0 \alpha_t^* \alpha_t^* \beta_s^* E_i^2}{16\pi\,(d+2a)^3}\right) \qquad (13)$$

Where $\beta_s = (\epsilon_s - 1)/(\epsilon_s + 1)$ is sample reflection coefficient, with $\epsilon_s$ the complex dielectric function of the tip. Effective tip polarizability $\alpha_t = \xi V_t (\epsilon_t - 1)/(\epsilon_t + 2)$, with $V_t$ the effective tip volume, $\xi$ is tip field enhancement factor when tip is far from surface, and $\epsilon_t$ is the complex dielectric function of tip. We have a similar expression for $\alpha_s$ with the parameters for sample replacing those of tip, except that $\xi = 1$ in the latter case. Based on our numerical simulations for the gold coated tip, $\xi \sim 10$. The last term in Eq. 13 is the most dominant term.

Equation 13 was evaluated with the appropriate Lorentzian dielectric functions for PMMA and dielectric constants for the tip[35,36]. Fig. 6d compares the experimental point spectrum (diamond symbol) with theory (solid line) showing excellent agreement. For the PiFM approach curve, $f_1$ was mechanically tuned to get the maximum PiFM signal for each setpoint (as shown in Fig. 6c). The tracked peaks were normalized to the corresponding $A_2$. Figure 6b shows the PiFM signal and the corresponding $A_2$ signal as a function of tip-sample distance. We see that the PiFM signal is measurable up to 18 nm from sample surface. Figure 6e shows a fit to the experimental approach curve using Eq. 13. The data fits a $1/d^3$ dependence as expected up to a tip-sample distance of 5nm. The PiFM signal, Fig. 6b, shows a characteristic feature at molecular resonance; the signal increases monotonically as the tip approaches the sample but then decreases at the inflection point at 4 nm. This damping feature has also been observed by X. G. Xu et .al.[37] in measurements of near field approach curves in sSNOM and was modeled by introducing a distance dependent phase shift into the tip reflection coefficient; they attributed scattering loss in sSNOM to dissipation due to optical re-radiation from the tip.

The inflexion point can be incorporated into our model by adding an additional complex distance -jb to the distance dependence[38,39].

$$<P_t> \sim \frac{1}{2}[\text{Im}(\omega_o \alpha_t^*)E_i^2 + \text{Im}\left(\frac{\omega_o \alpha_t^* \alpha_s^* E_i^2}{2\pi[(d+2a-jb)]^3}\right) + \text{Im}\left(\frac{\omega_o \alpha_t^* \alpha_s E_i^2}{2\pi[(d+2a-jb)]^3}\right) +$$

$$\text{Im}\left(\frac{\omega_o \alpha_t^* \alpha_t^* \beta_s^* E_i^2}{16\pi[(d+2a-jb)]^3}\right)] \quad (14)$$

While the overall force is attractive, this modification generates an additional distance dependent term varying as $\frac{-1}{(d+2a)^4}$. This distance dependence has been attributed to electron-hole pair absorption at the surface of the conductor[38,39]. Similar distance dependences have been predicted in models of the interaction energy of an atomic dipole oscillating close to a conducting sphere[40] (see Eq. 39 of reference 40). The modified Eq. 14 was evaluated with tip radius $a_t$ = 12 nm, sample radius $a_s$ = 5nm and the fitting parameter b = 1.7 nm. This modification generated an excellent fit between our model and the measured PiFM approach curve (Fig. 6e solid line).

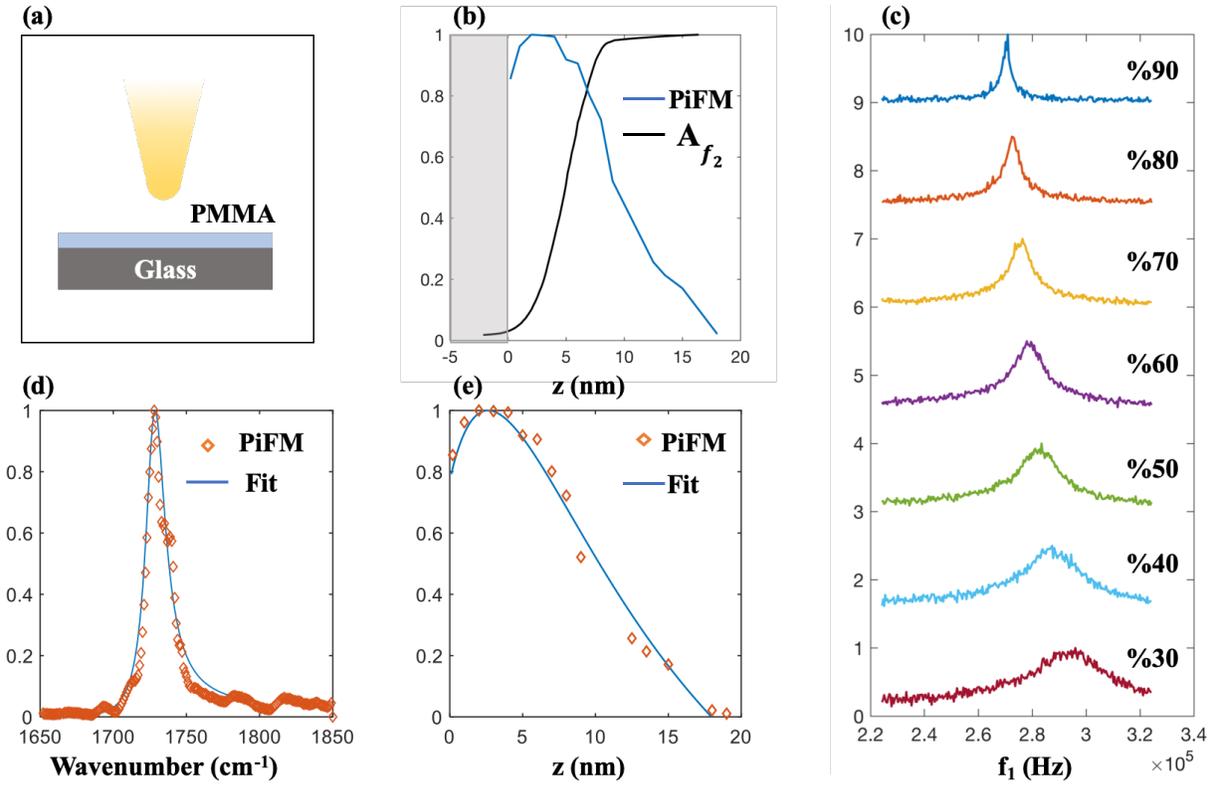

**Fig. 6** Experimental and theoritical results of opto-mechanical damping. **a** schematic of tip and sample. **b** amplitude $A_2$ measured at $f_2$ (black line) and PiFM signal measured at $f_1$ (blue line) as a function of tip-sample distance. The shift in the peak value of $f_1$ as a function of tip-sample distance was tracked in **c** and normalized by the correspondent amplitude measured at $f_2$. **d** and **e** are comparisons of experiement (diamond) and theory Eq. 14 (solid line) for point spectrum and approach curve respectivily.

I. Discussion

Mechanical dissipation induced by tip-sample interaction has already been studied in several AFM modalities in the 100KHz to MHz frequency range. Tapping mode is a prominent example for probing sample mechanical properties – the softer the sample the higher the mechanical loss. The observed energy loss in tapping mode is attributed to adhesion hysteresis. In this case, tip must permanently or temporally be contacting the sample. Noncontact mode has also been used to measure long-range dissipative interactions in doped semiconductors mediated by a vibrating-charged tip close to the surface. Near-field damping due to charge fluctuations in a dielectric sample, (polymer film), have been also been measured using non-contact AFM.

To our knowledge, opto-mechanical dissipation at optical frequencies has not been observed or discussed in the context of tip-based spectroscopy techniques; namely adsorbed molecules excited at one of its vibrational resonances exerting damping force on the tip leading to mechanical dissipation (Fig. 5).

II. Conclusion:

By tracking the $1^{st}$ mechanical eigenmode as a function of incident wavenumber, tip-sample optical interaction was shown to be attractive in nature. Maximum frequency shift was observed when sample was excited at its molecular vibrational resonance. Our observations clearly prove that short-range thermal expansion or long-range acoustics forces have negligible effect on PiFM signal. Thermal expansion of sample was successfully simulated by vibrating a calibrated PZT crystal. Noncontact-AFM sensitivity to thermal expansion was experimentally determined to be 1.53 µV/pm. PiFM sensitivity to monolayer sample (4-MBT) spectroscopy was experimentally demonstrated. Both PZT and monolayer experiments demonstrated that the PiFM signal could not be due to long-range thermal expansion mediated by van der Waal forces. In standard PiFM side-band mixing mode, PiFM signal contrast is dominated by the modulation of the *effective damping constant* (i.e. AM modulation of the cantilever resonance $f_2$) and not by modulation of the *dynamic effective spring* constant (i.e. FM modulation of cantilever second eigenmode). PiFM spectroscopic contrast for mid-IR vibrational resonance of molecules was experimentally demonstrated to be due to opto-mechanical damping detected through a change in the effective cantilever damping constant. Finally, our theoretical model shows excellent agreement with experimental results.

**References**

1. Wickramasinghe, H. K. & Williams, C. C. Apertureless near field optical microscope. US Patent # 4,947,034 (1990).
2. Zenhausern, F., O'Boyle, M. P. & Wickramasinghe, H. K. Apertureless near-field optical microscope. *Appl. Phys. Lett.* **65**, 1623–1625 (1994).
3. Knoll, B. & Keilmann, F. Near-field probing of vibrational absorption for chemical microscopy. *Nature* **399**, 134–137 (1999).

4. Zenhausern, F., Martin, Y. & Wickramasinghe, H. K. Scanning Interferometric Apertureless Microscopy: Optical Imaging at 10 Angstrom Resolution. *Science* **269**, 1083–1085 (1995).

5. Dazzi, A., Glotin, F. & Carminati, R. Theory of infrared nanospectroscopy by photothermal induced resonance. *J. Appl. Phys.* **107**, 124519 (2010).

6. Dazzi, A., Prazeres, R., Glotin, F. & Ortega, J. M. Local infrared microspectroscopy with subwavelength spatial resolution with an atomic force microscope tip used as a photothermal sensor. *Opt. Lett.* **30**, 2388–2390 (2005).

7. Wang, L. *et al.* Nanoscale simultaneous chemical and mechanical imaging via peak force infrared microscopy. *Sci. Adv.* **3**, e1700255 (2017).

8. Rajapaksa, I., Uenal, K. & Wickramasinghe, H. K. Image force microscopy of molecular resonance: A microscope principle. *Appl. Phys. Lett.* **97**, 073121 (2010).

9. Tumkur, T. *et al.* Wavelength-Dependent Optical Force Imaging of Bimetallic Al–Au Heterodimers. *Nano Lett.* **18**, 2040–2046 (2018).

10. Jahng, J. *et al.* Visualizing surface plasmon polaritons by their gradient force. *Opt. Lett.* **40**, 5058–5061 (2015).

11. Zeng, J. *et al.* Sharply Focused Azimuthally Polarized Beams with Magnetic Dominance: Near-Field Characterization at Nanoscale by Photoinduced Force Microscopy. *ACS Photonics* **5**, 390–397 (2018).

12. Tumkur, T. U. *et al.* Photoinduced Force Mapping of Plasmonic Nanostructures. *Nano Lett.* **16**, 7942–7949 (2016).

13. Rajaei, M., Almajhadi, M. A., Zeng, J. & Wickramasinghe, H. K. Near-field nanoprobing using Si tip-Au nanoparticle photoinduced force microscopy with 120:1 signal-to-noise ratio, sub-6-nm resolution. *Opt. Express* **26**, 26365–26376 (2018).

14. Huang, F., Ananth Tamma, V., Mardy, Z., Burdett, J. & Kumar Wickramasinghe, H. Imaging Nanoscale Electromagnetic Near-Field Distributions Using Optical Forces. *Sci. Rep.* **5**, 10610 (2015).

15. Rajapaksa, I. & Kumar Wickramasinghe, H. Raman spectroscopy and microscopy based on mechanical force detection. *Appl. Phys. Lett.* **99**, 161103 (2011).

16. Tamma, V. A., Beecher, L. M., Shumaker-Parry, J. S. & Wickramasinghe, H. K. Detecting stimulated Raman responses of molecules in plasmonic gap using photon induced forces. *Opt. Express* **26**, 31439–31453 (2018).

17. Tamma, V. A., Huang, F., Nowak, D. & Kumar Wickramasinghe, H. Stimulated Raman spectroscopy and nanoscopy of molecules using near field photon induced forces without resonant electronic enhancement gain. *Appl. Phys. Lett.* **108**, 233107 (2016).

18. Jahng, J. *et al.* Ultrafast pump-probe force microscopy with nanoscale resolution. *Appl. Phys. Lett.* **106**, 083113 (2015).

19. Gu, K. L. *et al.* Nanoscale Domain Imaging of All-Polymer Organic Solar Cells by Photo-Induced Force Microscopy. *ACS Nano* **12**, 1473–1481 (2018).

20. Nowak, D. *et al.* Nanoscale chemical imaging by photoinduced force microscopy. *Sci. Adv.* **2**, e1501571 (2016).

21. Huang, Y. *et al.* Spectroscopic Nanoimaging of All-Semiconductor Plasmonic Gratings Using Photoinduced Force and Scattering Type Nanoscopy. *ACS Photonics* **5**, 4352–4359 (2018).

22. Ladani, F. T. & Potma, E. O. Dyadic Green's function formalism for photoinduced forces in tip-sample nanojunctions. *Phys. Rev. B* **95**, 205440 (2017).

23. Almajhadi, M. & Wickramasinghe, H. K. Contrast and imaging performance in photo induced force microscopy. *Opt. Express* **25**, 26923–26938 (2017).

24. Yang, H. U. & Raschke, M. B. Resonant optical gradient force interaction for nano-imaging and -spectroscopy. *New J. Phys.* **18**, 053042 (2016).

25. Jahng, J., Potma, E. O. & Lee, E. S. Tip-Enhanced Thermal Expansion Force for Nanoscale Chemical Imaging and Spectroscopy in Photoinduced Force Microscopy. *Anal. Chem.* **90**, 11054–11061 (2018).

26. Garcia, R. & Herruzo, E. T. The emergence of multifrequency force microscopy. *Nat. Nanotechnol.* **7**, 217–226 (2012).

27. Murdick, R. A. *et al.* Photoinduced force microscopy: A technique for hyperspectral nanochemical mapping. *Jpn. J. Appl. Phys.* **56**, 08LA04 (2017).

28. Tam, A. C. Applications of photoacoustic sensing techniques. *Rev. Mod. Phys.* **58**, 381–431 (1986).

29. Lu, F., Jin, M. & Belkin, M. A. Tip-enhanced infrared nanospectroscopy via molecular expansion force detection. *Nat. Photonics* **8**, 307–312 (2014).

30. Jin, M. & Belkin, M. A. Infrared Vibrational Spectroscopy of Functionalized Atomic Force Microscope Probes using Resonantly Enhanced Infrared Photoexpansion Nanospectroscopy. *Small Methods* **0**, 1900018

31. Jahng, J., Kim, B., Lee, E. S. & Potma, E. O. Quantitative analysis of sideband coupling in photoinduced force microscopy. *Phys. Rev. B* **94**, 195407 (2016).

32. Anczykowski, B., Gotsmann, B., Fuchs, H., Cleveland, J. P. & Elings, V. B. How to measure energy dissipation in dynamic mode atomic force microscopy. *Appl. Surf. Sci.* **140**, 376–382 (1999).

33. Novotny, L. & Hecht, B. Principles of Nano-Optics by Lukas Novotny. *Cambridge Core* (2006).

34. Cvitkovic, A., Ocelic, N. & Hillenbrand, R. Analytical model for quantitative prediction of material contrasts in scattering-type near-field optical microscopy. *Opt. Express* **15**, 8550–8565 (2007).

35. Olmon, R. L. *et al.* Optical dielectric function of gold. *Phys. Rev. B* **86**, 235147 (2012).

36. Zolotarev, V. M., Volchek, B. Z. & Vlasova, E. N. Optical constants of industrial polymers in the IR region. *Opt. Spectrosc.* **101**, 716–723 (2006).

37. Wang, H., Wang, L., Jakob, D. S. & Xu, X. G. Tomographic and multimodal scattering-type scanning near-field optical microscopy with peak force tapping mode. *Nat. Commun.* **9**, 1–11 (2018).

38. Persson, B. N. J. Theory of the damping of excited molecules located above a metal surface. *J. Phys. C Solid State Phys.* **11**, 4251–4269 (1978).

39. Persson, B. N. J. & Lang, N. D. Electron-hole-pair quenching of excited states near a metal. *Phys. Rev. B* **26**, 5409–5415 (1982).

40. de Melo e Souza, R., Kort-Kamp, W. J. M., Sigaud, C. & Farina, C. Image method in the calculation of the van der Waals force between an atom and a conducting surface. *Am. J. Phys.* **81**, 366–376 (2013).

**Acknowledgements**

The authors thank Thomas R. Albrecht and Sung Park for stimulating discussions, and Derek Nowak for technical support on PiFM. The experimental and theoretical collaboration is supported by Chemistry at the Space-Time Limit (CaSTL) under Grant No. CHE-1414466 and Saudi Arabian Cultural Mission (SACM).

**Author Contributions**

M. A. A. conducted experiments. M.A.A and H.K.W interpreted the results. S. M. A. U. built interferometer setup and performed piezo calibration measurements. H. K. W. conceived and supervised the project. M. A. A. and H. K. W. wrote the manuscript.